\documentclass[aps,prl,amsmath,amssymb,reprint,superscriptaddress]{revtex4-1}
\usepackage{graphicx}
\usepackage{amsmath}
\usepackage{dcolumn}
\usepackage{bm}
\usepackage{bbold}
\usepackage{color}
\usepackage{amsfonts}
\usepackage{amssymb}
\usepackage{mathrsfs}
\usepackage{tabularx}
\usepackage{braket}
\usepackage{mathtools}
\usepackage{soul}			

\usepackage{caption}
\usepackage{subcaption}

\begin{document} 
\title{Single Photon Level Study of Microwave Properties of Lithium Niobate at milli-Kelvin Temperatures}
\author{Maxim Goryachev}
\affiliation{ARC Centre of Excellence for Engineered Quantum Systems, School of Physics, University of Western Australia, 35 Stirling Highway, Crawley WA 6009, Australia}

\author{Nikita Kostylev}
\affiliation{ARC Centre of Excellence for Engineered Quantum Systems, School of Physics, University of Western Australia, 35 Stirling Highway, Crawley WA 6009, Australia}

\author{Michael E. Tobar}
\email{michael.tobar@uwa.edu.au}
\affiliation{ARC Centre of Excellence for Engineered Quantum Systems, School of Physics, University of Western Australia, 35 Stirling Highway, Crawley WA 6009, Australia}

\date{\today}


\begin{abstract}

Properties of doped and natural impurities in Lithium Niobate single crystals are studied using the Whispering Gallery Mode method at low temperatures as a function of magnetic field. The study reveals considerable coupling of microwave photon modes to the Fe$^{3+}$ spin ensemble in iron-doped and non-doped crystals. The $S=5/2$ structure of the Fe$^{3+}$ impurities demonstrate Zero Field Splittings of $11.21$ and $20.96$ GHz, significant asymmetry of the Zeeman lines and additional lines with anomalous $\text{g}$-factors of $1.37$ and $3.95$. Also, interactions between different transitions of the Fe$^{3+}$ ion is observed. An additional ion impurity ensemble with a splitting of about $1.7$ GHz is shown to couple to the dominating Fe$^{3+}$ spins and the effect on $Q$-factors of microwave photon modes due to the Fe$^{3+}$ ion ensemble is also demonstrated. Measurements down to less than one photon level are made with a loss tangent of order $10^{-5}$ determined. 

\end{abstract}

\maketitle

Quantum system technology is a rapidly growing field that promises a new generation of sensing, computing and data transfer. For the last couple of decades, a variety of approaches to quantum technology has been developed including superconducting qubits, trapped ions, ions-in-solids, optomechanics, etc. All of these directions have their advantages and disadvantages. As a result, the concept of the Hybrid Quantum System (HQS) has been proposed and developed\cite{PhysRevA.87.052333}. The concept of HQS is not only supposed to overcome disadvantages of stand alone quantum technologies but also to couple them into a single network. An example of this is microwave-to-optical conversion with quantum efficiency\cite{PhysRevA.80.033810,PhysRevLett.109.130503,PhysRevLett.92.247902,Williamson:2014aa} that may be used to unify separated in space superconducting quantum processing units into a whole network via optical links. More generally, the problem of combination of optical and microwave subsystem is for the perspective quantum devices is important at the current stage of technology development. 

 The discussed optical-microwave coupling my be constructed using mechanical devices coupled simultaneously to optical and microwave cavities or with a spin ensemble exhibiting both microwave and optical transitions or using media with nontrivial properties in the both parts of the spectrum. The problem of the last two approaches is related to the absence of systems and materials that are very well investigated both in optical and microwave domains especially at low temperatures. This happens because typical working solutions in one field demonstrates considerable disadvantages in another and thus does not suits the combined system. To cover this gap between the two kingdoms, we investigate a known material with recognised nontrivial optical properties, namely Lithium Niobate (LiNbO$_3$)\cite{LiNb} in the microwave frequency range and at milli-Kelvin temperatures. Lithium Niobate is widely used for various optical applications including optical modulation, optical waveguides, high-$Q$ microdisc resonators\cite{Lin:2015aa,Furst:2010aa} exhibiting nonlinear phenomena\cite{appsL} including attempts to create microwave-to-optics conversion with quantum level efficiency\cite{PhysRevA.80.033810}. On the other hand, despite some work\cite{Morse:1994aa,Bravina:2004aa,Herzog:2008aa,Dhar:2013aa,Huang:2014aa}, mechanical, microwave and optical properties of this material at low temperatures have not been studied in detail.  
 
In this work, the microwave properties of LiNbO$_3$ single crystals are studied using the Whispering Gallery Mode (WGM) approach. This approach has been demonstrated to provide very accurate data for dielectric properties of low loss materials\cite{Krupka:1999aa} and ultra-sensitive spectroscopy of naturally occurring\cite{PhysRevB.88.224426,goryachev2013giant,Goryachev:2014aa,IGIYSO} and doped\cite{IGIYSO,PhysRevB.90.054409,YAGfarr} ion impurities. Depending on material losses, this technique can detect ion ensembles with concentrations down to few parts per billion at multiple frequencies in the X and K$_\text{u}$ bands ($5-25$ GHz). Additionally WGMs are a nice tool to observe various nonlinear phenomena related to dielectrics and spin ensembles such as four-wave mixing and masing\cite{Bourhill2013b,PhysRevLett.108.093902}. Advantages of the WGM technique are high filling factor (due to the fact that the microwave cavity is simultaneously a host for the spin ensemble) and high quality factors. 
The obtained results may be used to decide on whether a particular spin ensemble may be used for Quantum Electrodynamics (QED) experiments and what spectrum regions have to be avoided to minimise photon losses due to the two level system absorption.

The experiments are performed with two cylindrical single crystals: undoped LiNbO$_3$ and Iron doped LiNbO$_3$ (Fe$^{2+}$, Fe$^{3+}$ ions with 0.005 wt\%), with the c-axis of the crystal aligned with the z-axis of the cylinder (the crystal exhibits biaxial anisotropy).
Both crystals are $15$mm long and $15$mm diameter with 1mm hole in the centre for metallic support structure. Crystals are enclosed in a closed oxygen free copper structure with a two straight probe coupling antennas. The whole system is placed in the centre of a 7T superconducting magnet and thermally grounded to a 20mK stage of a dilution cryogenerator. The spectroscopy is performed via network analysis with the room temperature signal attenuated by a series of cryogenic attenuators ($-40$dB at different cryocooler stages). The output signal passess through low noise amplifiers both at 4K and at room temperature. The cavity and the cryogenic amplifier are separated by a milli-Kelvin circulator to prevent the back action noise from the 4k stage of the amplifier. The setup is similar to previous QED experiments with WGM and other microwave cavities\cite{Goryachev:2014aa,Goryachev:2014ab}.

\begin{figure}[t!]
			\includegraphics[width=0.5\textwidth]{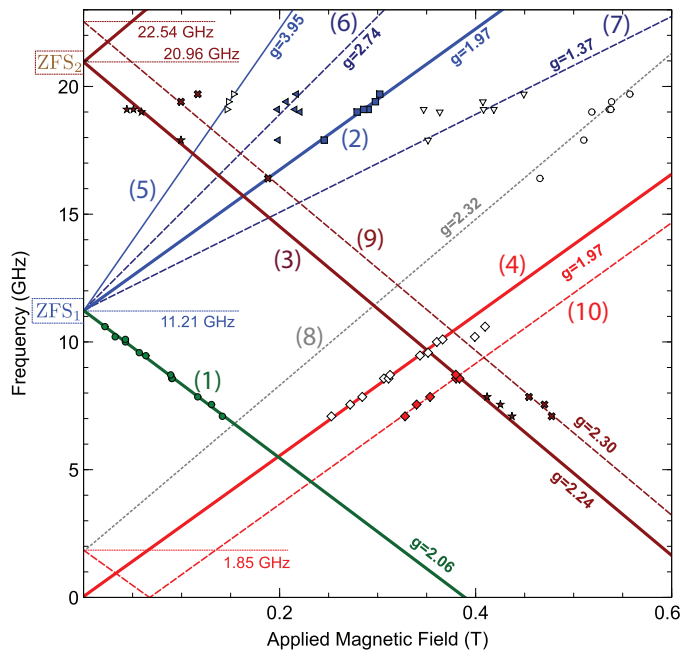}
	\caption{Spectroscopy results for the Fe:LiNbO$_3$ crystal. Point objects mark experimentally observed points of interaction. Lines demonstrate their interpretation in terms of the Zeeman tuning of transition in ion ensembles. }
	\label{spec1}
\end{figure}

\begin{figure}[t!]
			\includegraphics[width=0.5\textwidth]{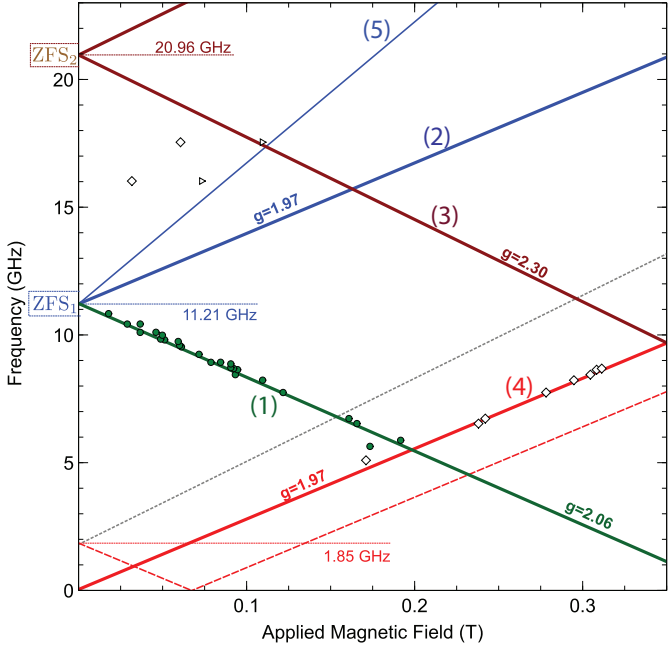}
	\caption{Spectroscopy results for the undraped LiNbO$_3$ crystal. Lines are fits for the Fe:LiNbO$_3$ crystal}
	\label{spec2}
\end{figure}

Resonance frequencies of WGMs of cylindrical crystals are set by the cavity dimensions and dielectric and magnetic properties of the material. In the case of paramagnetic spin ensemble impurities within the crystal, the external magnetic field changes the energy level splitting due to the Zeeman effect. When the splitting frequency coincides with the resonance frequency of one of the system WGMs, the latter exhibits some frequency and line width deviation. So, in order to reveal the ion spin level structure, we construct a map of interactions where each point represent an interaction frequency and the corresponding magnetic fields. The interaction maps for the doped and undoped Lithium Niobate are shown in Fig.~\ref{spec1} and \ref{spec2} correspondingly. 

The experimental data shown in Fig.~\ref{spec1} reveals a Zero Field Splitting (ZFS) structure (ZFS$_1$ and ZFS$_2$) with Zeeman lines (1), (2), (3), (4) and (5) that may be interpreted as the transition structure between energy levels of Fe$^{3+}$ ions\cite{Malovichko:1993aa,Keeble:2005aa}. Here line (5) is a two photon transition ($\Delta S_m = 2$) with the $\text{g}$-factor twice of that of the line (2). By fitting two ZFS ($11.21$GHz and $20.96$GHz), one finds the spin parameter $D=5.33$~GHz that approximately corresponds to that predicted before\cite{Keeble:2005aa}. 

Despite good agreement of ZFS parameters with previous results, the spin ensemble also demonstrates a few distinct features that have not been observed at higher temperatures. Firstly, $\text{g}$-factors for Zeeman lines with positive and negative directions have different sign corrections to the electron $g$-factor in vacuum making the splitting slightly asymmetric. This property has been also observed for natural impurities in quartz\cite{goryachev2013giant}. Secondly, besides two Zeeman lines (2) and (5) representing expected one and two ion transitions, the map of interactions demonstrates existence of two more lines with the same ZFS (lines (6) and (7)). Their anomalous $\text{g}$-factors ($1.37$ and $3.95$) are considerably different from the vacuum electron spin $\text{g}$-factor with one being twice the other. The origin of these Zeeman lines cannot be explained by the traditional spin-Hamiltonian of Fe$^{3+}$ ions in solids. Thirdly, Zeeman lines (8), (9) and (10) with ZFS of $1.85$~GHz and $22.54$~GHz do not correspond to any transition of Fe$^{3+}$ $S=5/2$ structure and are most likely another spin ensemble. It can be noted that line (10) is related to line (4) by a similar amount ($1.85$~GHz) of zero field frequency shift as line (9) to line (3) ($1.55$~GHz) also corresponding to the ZFS of line (8). This might be explained by coupling of the Fe$^{3+}$ to some other ion species with the corresponding ZFS. Similar phenomenon for Fe$^{3+}$ and V$^{2+}$ have be proposed to explain extra hyperfine structure in sapphire\cite{PhysRevB.88.224426}. Presence of other Iron Group ions such as Cr$^{3+}$\cite{Yeom:1993aa}, Mn$^{2+}$\cite{MSrev91} and Cu$^{2+}$\cite{MWrev91} in LiNbO$_3$ at room temperature have been studied at room temperature. On the other hand, Iron ions themselves can come in different cites centres giving rise to much smaller crystal field parameters: $b_2^{0}=1.5$GHz and $2.1$GHz\cite{Malovichko:1993aa} and $b_2^{0}=0.46$GHz and $0.6$GHz\cite{bd74}.

Spectroscopy results for the undoped crystal clearly demonstrated only two transition lines (Fig.~\ref{spec1}). These transitions correspond to lines (1) and (4) of the doped crystal. This fact demonstrates some natural abundance of Iron in LiNbO$_3$. 

\begin{figure}[t!]
			\includegraphics[width=0.5\textwidth]{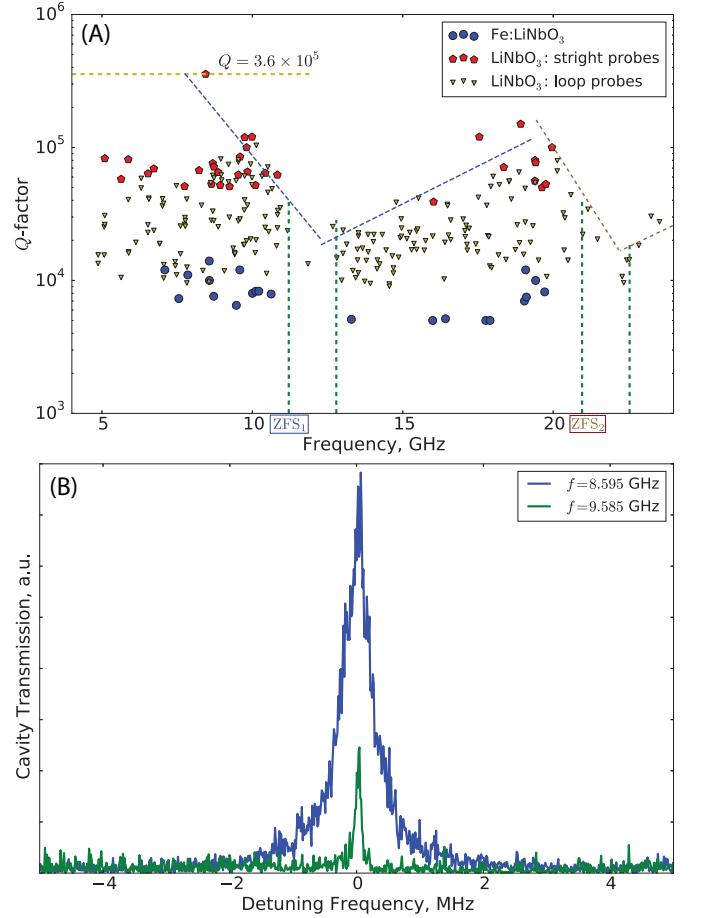}
	\caption{(A) $Q$-factors of WGMs used for impurity spectroscopy. (B) Mode shapes of two WGMs measured at less than one stored photon.}
	\label{qfact}
\end{figure}

Relation between paramagnetic impurities and microwave losses is well known\cite{Creedon:2011wk,Hartnett:2001aa}. This relation may be observed via quality factors of microwave WGMs. Fig.~\ref{qfact}, (A) displays $Q$-factors for WGMs for both doped and undoped crystals over the available frequency band. The plot shows that losses in the doped sample are about an order of magnitude exceed that in the undraped crystal.
The other important detail is relation between the fist ZFS and the observed spectrum: the mode density drops significantly immediately above ZFS$_1$ making a gap in the mode spectrum. Measurements of the cavity transmission as a function of the driving power demonstrated no clear dependence on the number of stored cavity photons. Fig.~\ref{qfact}, (B) demonstrates the cavity transmission near two WGMs measured with less than one stored photon on average\cite{Creedon:2011wk}. 

\begin{figure}[t!]
			\includegraphics[width=0.5\textwidth]{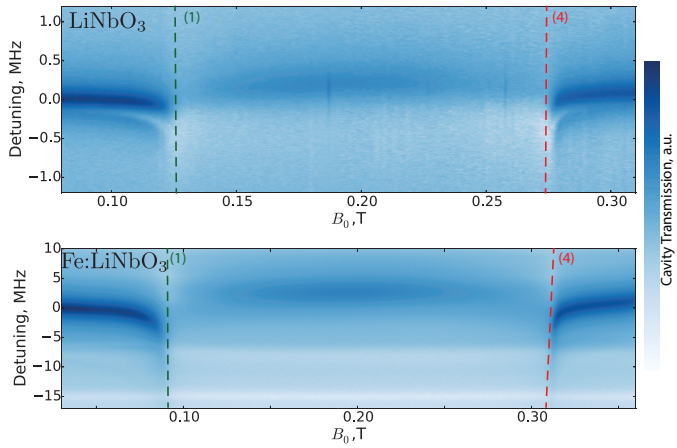}
	\caption{Frequency response of doped and undraped crystals as a function of external magnetic field $B_0$ in close proximity between spin transitions (1) and (4).}
	\label{inter}
\end{figure}

\begin{figure}[t!]
			\includegraphics[width=0.5\textwidth]{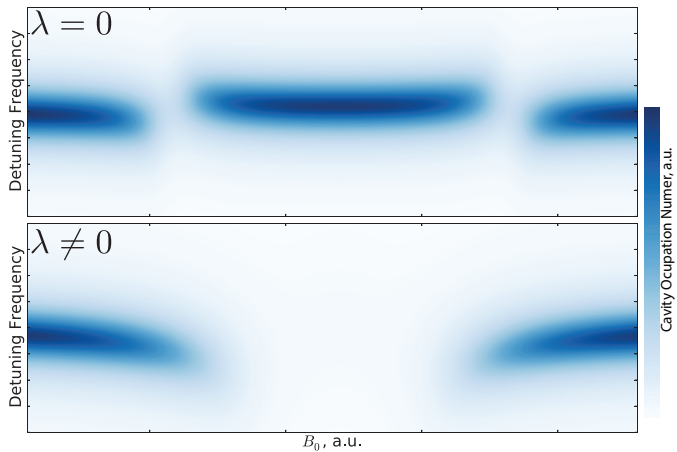}
	\caption{Modeling of the cavity mode interaction with two ion transitions corresponding to $\lambda=0$ of different ions, $\lambda\neq0$ of the same ion.}
	\label{simul}
\end{figure}

Another important feature of Fe$^{3+}$ ions in Lithium Niobite is strong influence of different transitions on each other. The property that can be important for QED type experiments. In previous experiments with low doped crystals\cite{PhysRevB.88.224426,goryachev2013giant,Goryachev:2014aa,IGIYSO}, each interaction is well approximated by an Avoided Level Crossing (ALC between an ensemble of Two Level System (TLS) and a Harmonic Oscillator (HO). For the Iron ensemble in Lithium Niobite, the situation is different as depicted in Fig.~\ref{inter} where interaction lines (1) and (4) display intercoupling resulting in ALC suppression. This situation can be modelled by a cavity mode $a^\dagger a$ interacting with two TLS ensembles with non-negligible intercoupling:
\begin{multline}
	\label{B006SFa}
	\displaystyle  H = \omega_ca^\dagger a+\sum_i\big[\omega_{a1}\sigma^z_{1i} + \omega_{a2}\sigma^z_{2i} +\\
	\displaystyle+\lambda(\sigma^+_{1i}\sigma^-_{2i}+\sigma^+_{2i}\sigma^-_{1i})+g_i(\sigma^+_{1i}a+a^\dagger\sigma^-_{1i}+\sigma^+_{2i}a+a^\dagger\sigma^-_{2i})\big],
\end{multline}
where $\sigma^+_{ni}$, $\sigma^+_{ni}$ and $\sigma^+_{ni}$ are the usual spin operators for the $n$th transition of the $i$th ion, $\lambda$ is inter-transition coupling. The frequency response of this system including its couplings to the environment is simulated using QuTiP\cite{Johansson20121760,Johansson20131234}. The result shown in Fig.~\ref{simul} displays the cavity occupation number as a function of the pump frequency and external magnetic field in two cases: independent ion transitions ($\lambda = 0$) and the same ion transitions ($\lambda\neq0$). The simulation clearly demonstrates that the interaction with the transitions of the same ion may lead to disappearance of ALCs even if they are separated by more than the ion-cavity coupling strength.

In summary, we investigated properties of Iron doped and undoped Lithium Niobate at milli-Kelvin temperatures with WGMs. The microwave spectroscopy confirms existence of Fe$^{3+}$ with ZFSs of $11.21$ and $20.96$GHz in both samples. Additional Zeeman lines with anomalous $\text{g}$-factors of $1.37$ and $3.95$ are suggested. 
Additional structure with a characteristic splitting of $1.85-1.55$~GHz suggests existence of another coupled spin ensemble. Influence of the impurity ensemble is revealed via $Q$-factors and density of microwave modes. 

%

\end{document}